\documentclass[%
 reprint,
superscriptaddress,
 amsmath,amssymb,
 aps,
floatfix,
]{revtex4-2}

\usepackage{graphicx}
\usepackage{dcolumn}
\usepackage{bm}
\usepackage{braket}
\usepackage{siunitx}
\usepackage[T1]{fontenc}

\usepackage{hyperref}
\hypersetup{
    colorlinks=true,
    linkcolor=magenta,
    citecolor=magenta,
    filecolor=magenta,      
    urlcolor=magenta,
}

\begin{document}

\title{Nonlinear response and crosstalk of electrically driven silicon spin qubits}

\author{Brennan~Undseth}
 \thanks{These authors contributed equally to this work;}
\author{Xiao~Xue}
 \thanks{These authors contributed equally to this work;}
\author{Mohammad~Mehmandoost}
\author{Maximilian~Russ}
\affiliation{QuTech and Kavli Institute of Nanoscience, Delft University of Technology, Lorentzweg 1, 2628 CJ Delft, The Netherlands}
\author{Pieter~T.~Eendebak}
\author{Nodar~Samkharadze}
\author{Amir~Sammak}
\affiliation{QuTech and Netherlands Organization for Applied Scientific Research (TNO), Stieltjesweg 1, 2628 CK Delft, Netherlands}
\author{Viatcheslav~V.~Dobrovitski}
\author{Giordano~Scappucci}
\author{Lieven~M.~K.~Vandersypen}
\affiliation{QuTech and Kavli Institute of Nanoscience, Delft University of Technology, Lorentzweg 1, 2628 CJ Delft, The Netherlands}

%\date{\today}

\begin{abstract}

Micromagnet-based electric dipole spin resonance (EDSR) offers an attractive path for the near-term scaling of dense arrays of silicon spin qubits in gate-defined quantum dots while maintaining long coherence times and high control fidelities. However, accurately controlling dense arrays of qubits using a multiplexed drive will require an understanding of the crosstalk mechanisms that may reduce operational fidelity. We identify a novel crosstalk mechanism whereby the Rabi frequency of a driven qubit is drastically changed when the drive of an adjacent qubit is turned on. These observations raise important considerations for scaling single-qubit control.

\end{abstract}

\maketitle % comment for word count

\section{Introduction}
%% Start Introduction %%
Electric dipole spin resonance (EDSR) is a key ingredient for the all-electrical control of single-electron spin qubits in silicon quantum dots \cite{Vandersypen_2017}. While some approaches are able to utilize the weak intrinsic spin-orbit coupling (SOC) of silicon \cite{Corna_2018, Huang_2017}, the placement of an on-chip micromagnet has proven especially effective for gate-based quantum dots in both Si/SiGe \cite{Kawakami_2014,Croot_2020_FloppingMode} and Si-MOS \cite{Leon_2020} platforms, with single-qubit gate fidelities exceeding 99.9\% having been demonstrated \cite{Yoneda_2017}. Furthermore, electron spins in dense arrays can be made addressable by engineering an appropriate local magnetic field gradient within a stronger external field \cite{Philips_2022_6qubit}. This makes micromagnet-based EDSR attractive for the near-term scaling of spin qubit processors.

In the original description of micromagnet-based EDSR, an ac electric field pushes a harmonically confined electron back and forth in a constant magnetic field gradient, such that the spin is effectively acted upon by an ac magnetic field as in conventional electron spin resonance (ESR) \cite{Tokura_2006, Pioro_Ladri_re_2008}. An array of spectrally-separated spins can ideally be controlled via a single, multiplexed driving field containing a linear combination of frequencies addressing individual qubits. Rabi's formula implies that the qubit dynamics are only slightly affected by off-resonance tones such that crosstalk can be accounted for systematically to maintain high fidelity \cite{Heinz_2021_crosstalkA}.

Substantial effort has been placed on detecting and modelling crosstalk in superconducting and trapped-ion systems \cite{Sarovar_2020}, but the identification of crosstalk mechanisms in semiconductor quantum dot devices is only beginning to receive attention as these platforms mature into the multi-qubit era \cite{Xue_2019_Benchmarking,Federe_2021_GaAsArray,Lawrie_2021_simultaneous}. Given that high qubit density is one of the known advantages of semiconductor quantum processors, maintaining high-fidelity operation with small qubit pitch in the presence of crosstalk is an essential hurdle to overcome.

In this article, we measure the nonlinear Rabi frequency scaling of two single-electron spin qubits controlled via EDSR in a $^{28}$Si/SiGe double-dot device. The nonlinearity gives rise to a sizeable crosstalk effect when attempting to drive simultaneous single-qubit rotations, and we develop a simple phenomenological extension of silicon-based EDSR theory to relate our observations. Although the physical origin of the nonlinearity is not precisely known, we find that anharmonicity in the quantum dot confining potential cannot quantitatively explain our measurements. We therefore comment on other device physics, such as microwave-induced artefacts, that may contribute to the crosstalk mechanism. The insights made here will help inform continued development of EDSR-enabled spin qubit devices, as well as raise important considerations for programming spin-based quantum processors in silicon.
%% End Introduction %%

%% Start Experiment %%
\begin{figure}[ht]
    \centering
    \includegraphics[width=\linewidth]{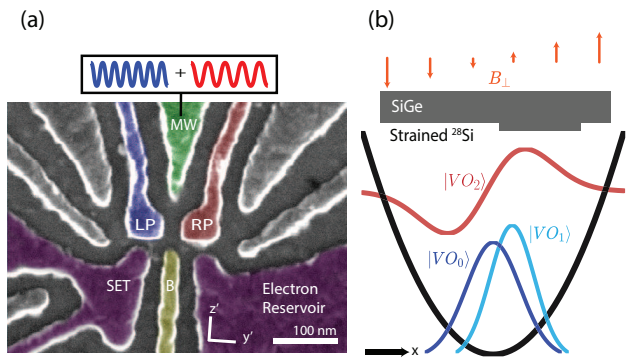}
    \caption{(a) False-coloured image of a device nominally equivalent to the one used in the experiment. Single-electron spin qubits Q1 and Q2 are confined under plunger gates ``LP'' and ``RP'' respectively, while a barrier gate ``B'' is used to control the tunnel coupling between the dots. Qubit states are read-out using energy-selective tunneling to the electron reservoir, with a single-electron transistor (SET) used to measure the corresponding change in charge-occupation. Microwave controls for both qubits are simultaneously applied to either the ``MW'' or ``B'' gate. (b) Illustration of wave function envelopes in a silicon quantum well. EDSR can be mediated by both the first excited orbit-like state $\ket{VO_2}$ as well as the first excited valley-like state $\ket{VO_1}$ as a consequence of interface-induced hybridization. Interface disorder here is represented as a rectangular ``atomic step'' for simplicity, but hybridization may also be a consequence of more detailed alloy disorder. In any case, a finite dipole transition element along with the micromagnet spin-orbit coupling enables electrically driven spin rotations.}
    \label{fig:1}
\end{figure}

\section{Methods}

To probe the behaviour of two spin qubits controlled via a frequency multiplexed drive, two quantum dots with single-electron occupancy are electrostatically accumulated in an isotopically purified $^{28}$Si/SiGe quantum well [Fig.~\ref{fig:1}(a)]. A cobalt micromagnet placed on top of the dot region becomes magnetized in the external field applied along the $z'$-axis, creating local transverse ($x'$-axis) and longitudinal ($z'$-axis) magnetic field gradients. The transverse gradient gives rise to a synthetic SOC, and the longitudinal gradient spectrally separates the Larmor frequencies of the two spins. The IQ-modulated electric drive necessary to control the spin states by EDSR is delivered via the gate ``MW'' or the gate ``B''. Further details of the fabrication, initialisation, control, and readout of the qubits can be found in \cite{Xue_2021_CMOS}.

\begin{figure}[ht]
    \centering
    \includegraphics[width=\linewidth]{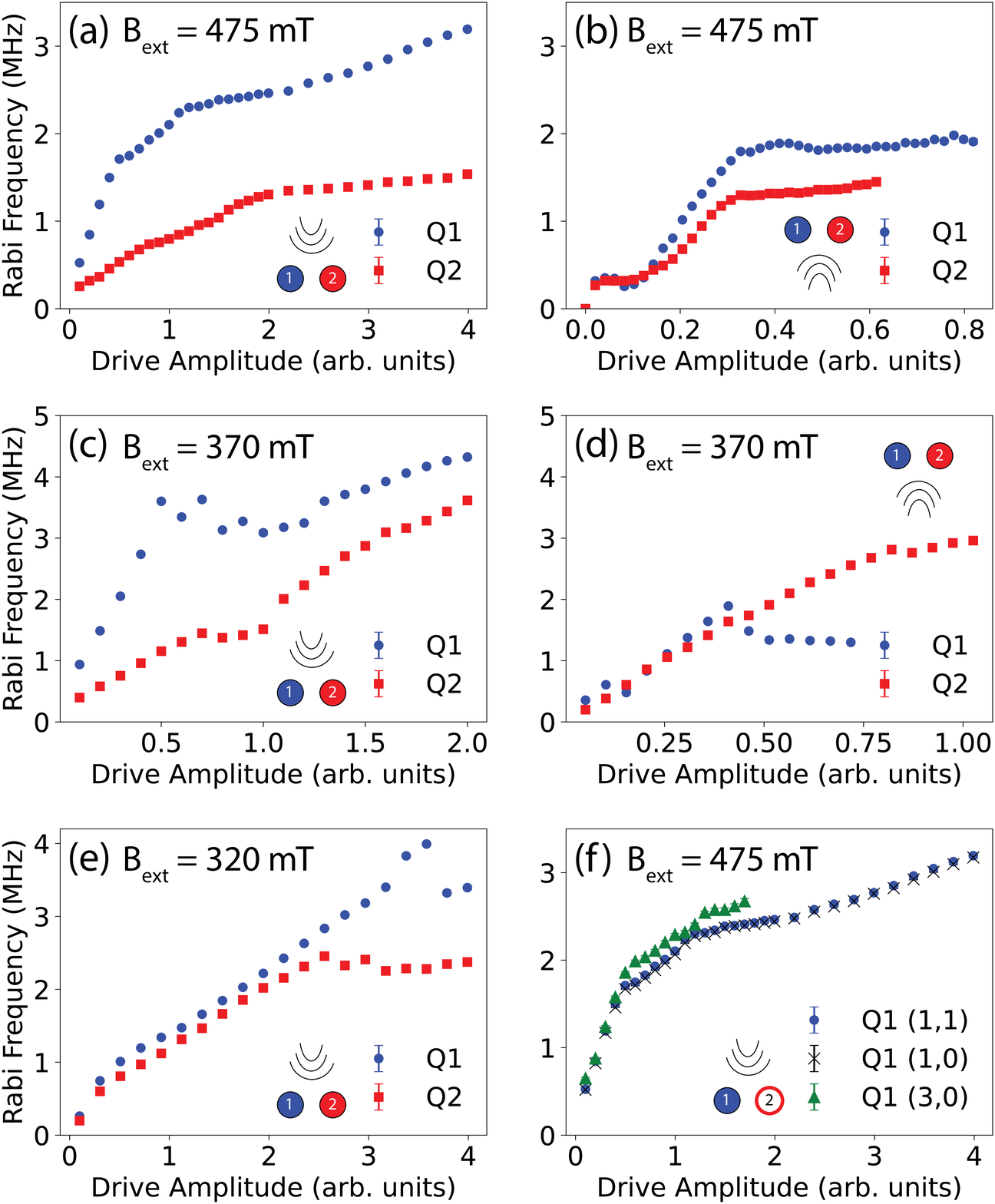}
    \caption{Rabi frequency scalings as a function of the applied resonant ac electric field amplitude. The external field is set to $B_\mathrm{ext}=$~\SI{475}{\milli\tesla} in (a),(b) and (f), $B_\mathrm{ext}=\SI{370}{\milli\tesla}$ in (c) and (d), and $B_\mathrm{ext}=$~\SI{320}{\milli\tesla} in (e). In (a), (c), (e), and (f) the qubits are driven using the ``MW'' gate as illustrated. In (b) and (d) the qubits are instead driven using the ``B'' gate. In (a-e) only a single qubit is driven at once in the (1,1) electron occupation regime, while the undriven qubit is left to idle. In (f) the Q1 Rabi scaling is compared in different charge states of the device. The horizontal axis is scaled such that 1 arbitrary unit (arb. unit) represents the same nominal drive amplitude delivered to the device by taking into account the room-temperature vector source power and all nominal attenuation in the signal paths.}
    \label{fig:2}
\end{figure}
In a gate-defined quantum dot in silicon, an electric field is able to couple spin-like qubit states via EDSR due to the spin-orbit coupling perturbing the pure spin eigenstates such that they become slightly hybridized with the electron charge states. For single-electron spin qubits in Si/SiGe, the charge states are themselves hybridized valley-orbit states owing to the nearly-degenerate conduction band valleys of strained silicon quantum wells \cite{Zwanenburg_2013}. EDSR may therefore be mediated by orbit-like or valley-like hybridized states which support a nonzero dipole transition element with the electron ground state, as illustrated in Fig.~\ref{fig:1}(b) \cite{Tokura_2006,huang2021fast}. In either case, a robust linear relationship between the amplitude of the driving field and the Rabi frequency of the spin qubit is expected (see Appendices B and C).

To drive on-resonance Rabi oscillations, we first use a Ramsey pulse sequence to accurately identify the relevant resonance frequencies of the two qubits, which range from \SI{11.89}{\giga\hertz} in a \SI{320}{\milli\tesla} external field to \SI{15.91}{\giga\hertz} in \SI{475}{\milli\tesla}. The corresponding drives are applied either to the ``MW'' or ``B'' gate, and the same driving frequency is used for all drive durations and amplitudes. A rectangular pulse with duration up to \SI{3}{\micro\second} is used, and the measured time-domain spin response is fit to a sinusoidal function $A\cos(2\pi f_\mathrm{Rabi}t+\phi)+C$ to extract the Rabi frequency.

\section{Results}

\subsection{Nonlinear Rabi Scaling}

We observe unexpected nonlinear Rabi frequency scaling when each spin is driven individually as shown in Fig.~\ref{fig:2}. In most cases, the linear Rabi frequency-drive amplitude scaling predicted from theory only holds for Rabi frequencies up to 1-\SI{2}{\mega\hertz}. The exact electric field driving amplitude is not known precisely, so a linear scale is used such that 1 unit of amplitude is approximately equivalent to a \SI{2}{\mega\hertz} Rabi frequency for Q1 in the configuration of Fig.~\ref{fig:2}(a). This nominal amplitude is used as a reference for other experiments, when the attenuation in each line can be used to estimate the power delivered to the device \footnote{Unknown impedance mismatches at, for example, wire bonds, make the on-chip ac electric field vary with frequency. Therefore the comparison between Rabi scaling trends at different magnetic fields is qualitative. Previous photon-assisted tunneling measurements on similar devices estimate the electric field amplitude to be on the order of \SI{1000}{\volt/\meter}, or equivalently a voltage amplitude of \SI{0.1}{\milli\volt} on the gate  \cite{Kawakami_thesis}. This estimate, along with a transverse magnetic gradient of order \SI{0.1}{\milli\tesla/\nano\meter} from simulations and a harmonic potential energy scale of \SI{1}{\milli\eV}, agrees with the frequency of Rabi oscillations we observe.}.

For each quantum dot, external magnetic field, and driving gate, the associated curve contains unique, but robustly reproducible, nonlinear characteristics qualitatively similar to \cite{Leon_2020}. These often appear as ``plateaus'' where the Rabi frequency apparently saturates, or only changes modestly, when the amplitude of the electric drive is adjusted. Increasing the driving amplitude does not always yield larger Rabi frequencies, nor is the visibility or quality of Rabi oscillations degraded in the plateau. In some experimental configurations, driving even more strongly in the nonlinear regime will lead to a sudden loss in qubit visibility. Decreased visibility and a diminished $T_2^\mathrm{Rabi}$ have been previously reported for fast EDSR in silicon \cite{Takeda_2016,Yoneda_2017}, and may be a result of population leakage to spin-orbit states outside of the qubit subspace. We speculate that the abrupt change in visibility may also occur from microwave heating interfering with the energy-selective readout used in the experiment.

In addition to the general Rabi saturation effect observed, each measured Rabi scaling exhibits distinct kinks. Note that the difference in scaling trends between adjacent spins has previously been observed \cite{Obata_2010} and may be attributed to differences in the local magnetic field gradient at each dot location. However, this does not explain the nonlinearity in the qubit response as the micromagnet gradient is nearly constant over the $\SI{100}{\nano\meter}$ pitch of the dots. From the distinct shapes of the Q1 and Q2 curves, it is apparent that the origin of the nonlinearity is particular to each qubit frequency and not a global phenomena as could be expected from a uniform distortion in the driving field. We also note that a drive-induced shift in the qubit's resonance frequency, which has previously been observed in EDSR experiments \cite{Takeda_2018_freqshift, Watson_2018, Philips_2022_6qubit}, is not a plausible cause of the nonlinear scaling since an off-resonant drive will result in faster oscillations, not slower \cite{Romhanyi_2015_EDSR}.

Next, we consider the possibility that the nonlinearity is due to the influence of the second qubit. However, upon removing the Q2 electron, there is no change in the Q1 Rabi scaling as shown in Fig.~\ref{fig:2}(f). Furthermore, the residual exchange interaction between the two qubits is measured to be below \SI{50}{\kilo\hertz}, indicating a very weak spin-spin interaction taking place. Repeating the experiment in the (3,0) regime produces the same initial linear trend, suggesting that in both the 1-electron and 3-electron modes the same dipole transition element, whether orbit-like or valley-like, is responsible for mediating EDSR. The nonlinear scaling regime is similarly shaped, but measurably different, suggesting that the root cause of the nonlinearity may be somewhat influenced by the quantum dot structure.

\begin{figure}[t]
    \centering
    \includegraphics[width=\linewidth]{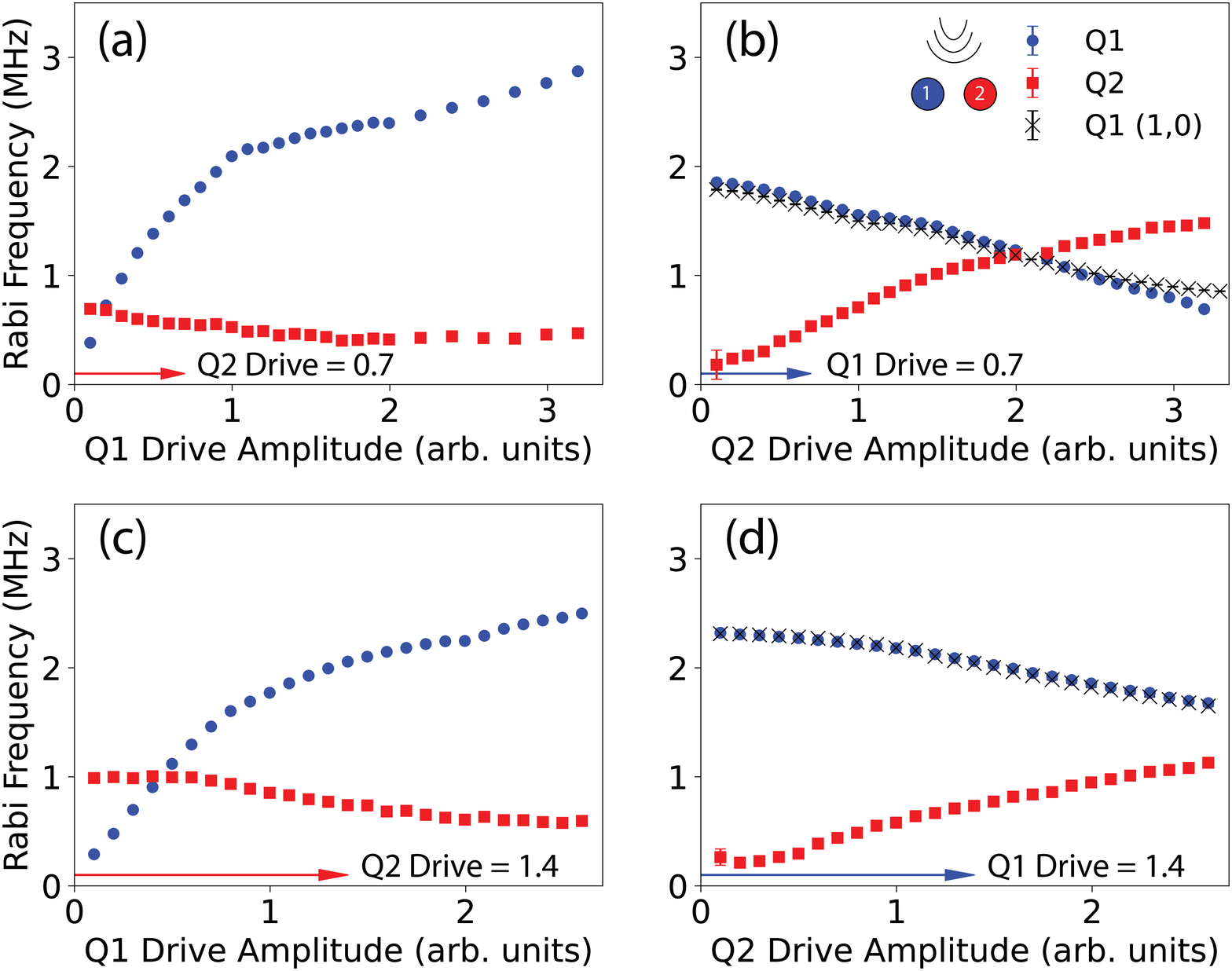}
    \caption{Crosstalk in single-qubit operation. The Rabi frequencies of both qubits are measured when a constant driving tone on-resonance with one qubit is present (shown in the bottom left of each panel) while a second tone resonant with the other qubit is swept in amplitude. In panels (a) and (b) the constant driving amplitude is half (0.7 arb. units) that in panels (c) and (d) (1.4 arb. units). Note that doubling the constant tone amplitude does not double the Rabi frequency, because the EDSR response is already nonlinear as shown in Fig.~\ref{fig:2}(a). In panels (b) and (d), the experiment repeated in the (1,0) regime gives nearly identical results as in the (1,1) regime. All experiments are carried out at $B_\mathrm{ext}=$~\SI{475}{\milli\tesla}, and the ``MW'' gate is used in all cases as indicated in the top right illustration.}
    \label{fig:3}
\end{figure}
\subsection{Crosstalk}

When both qubit driving tones are simultaneously applied to the ``MW'' gate, a large crosstalk effect occurs (Fig.~\ref{fig:3}). When a resonantly driven spin is also placed under the influence of an additional off-resonant drive, the additional ac field amplitude modifies the qubit response as to diminish the resonant spin-flip Rabi frequency. This effect has substantial consequences for high-fidelity logic gates which must be calibrated to a nanosecond-precision duration, because even a small unaccounted change in Rabi frequency would result in severe over- or under-rotations of qubit states. High-fidelity control can be maintained in a small device by operating gates serially \cite{Xue_2022_twoqubitgate,Noiri_2022_highfidelity,Mills_2022_highfidelity}, but this is an undesirable constraint for efficiently implementing quantum algorithms.

By comparing Fig.~\ref{fig:3}(a-b) with (c-d), it is clear that the Rabi frequency is more strongly modified when the resonant tone amplitude is smaller with respect to the off-resonant pulse amplitude. This implies that crosstalk would become more severe as single-qubit operations are more densely multiplexed. Directly adjusting microwave pulses for the unique response of each qubit may greatly increase the calibration overhead for larger qubit arrays, depending on the locality of the nonlinear response. We emphasize again that the crosstalk effect is not a consequence of the \textit{existence} of a nearby qubit, but rather is caused by the act of driving a second nearby qubit. This is illustrated in Fig.~\ref{fig:3}(b) and (d), where the Q1 behaviour is nearly identical in the case that the Q2 electron is removed from the double-dot region.
%% End Experiment %%

\section{Discussion}

%% Start Model %%
We now introduce a model Hamiltonian to survey in more depth the possible origin of the observed nonlinearity and crosstalk. Consider the following micromagnet-enabled EDSR Hamiltonian:
% This Hamiltonian is expressed with respect to the drive direction
% uncomment iffalse for word count
%\iffalse
\begin{equation}
    \label{eq:EDSR_Hamiltonian}
    H(t) = H_0 - \frac{E_Z}{2}\sigma_z + b'_{SL}\hat{x}\vec{n}\cdot\vec{\sigma} + E'_{ac}(t)\hat{x}.
\end{equation}
%\fi
%
\noindent $H_0$ describes the orbital and valley degrees of freedom of the charge state. $E_Z=g\mu_B B_\mathrm{tot}$ is the Zeeman splitting of the spin state, where $g\approx2$ is the g-factor in silicon, $\mu_B$ is the Bohr magneton, and $B_\mathrm{tot}$ is the total magnetic field along the $\sigma_z$ spin quantization axis.  $b'_{SL}=\frac{1}{2}g\mu_B|\vec{b}_{SL}|$ gives the strength of the SOC as a function of the magnitude of the magnetic field gradient $|\vec{b}_{SL}|$ along the driving axis ($x$). $\vec{n}=\begin{pmatrix} 0,\cos\theta,\sin\theta \end{pmatrix}^T$ characterizes the nature of the SOC, where $\theta$ gives the angle of the gradient with respect to the $\sigma_y$ spin quantization axis. The last term describes the electric drive $E'_{ac}(t) = e\sum_k E_{ac,k}\sin(\omega_k t)$ oriented along the $x$-axis.

EDSR is simplest to investigate in the case of harmonic confinement of the electron with effective mass $m^*$, such that $H_0=\hbar\omega_0(\hat{a}^\dagger\hat{a}+\frac{1}{2})$ with $\hat{a}^\dagger,\hat{a}$ being the quantum raising and lowering operators, $\hbar\omega_0$ giving the energy difference between orbital eigenstates, and $\hbar = h/2\pi$ as the reduced Planck's constant. The resulting Hamiltonian $H(t)$ can be analyzed perturbatively (see Appendix B) to find an on-resonance Rabi frequency of:

\begin{equation}
    \label{eq:Rabi_scaling}
    f_\mathrm{Rabi} = \frac{g\mu_B|\vec{b}_{SL}|\cos\theta e E_{ac}}{2hm^*\omega_0^2}
\end{equation}

\noindent and a drive-dependent resonance frequency shift of $\hbar\omega\propto -E_{ac}^2$ \cite{Romhanyi_2015_EDSR}. According to Equation~\ref{eq:Rabi_scaling}, dot-to-dot variations in EDSR sensitivity are expected as different qubit locations will experience different confinement strengths, magnetic field gradients, and electric driving angles. However, proportionality to the oscillating electric field amplitude is always expected from Eq.~\ref{eq:Rabi_scaling}.

Different linear scaling for small drives has been reported in both GaAs \cite{Obata_2010,Yoneda_2014_fastEDSR, Nakajima_2020} and Si \cite{Kawakami_2014, Takeda_2016,Yoneda_2017, Leon_2020}. The linear regime may extend from Rabi frequencies of only a few \SI{}{\mega\hertz} to tens of \SI{}{\mega\hertz} depending on the quantum dot environment, but a nonlinear regime can be identified when $f_\mathrm{Rabi}\ne B E_{ac}$ where $B$ is a scaling constant. Although smooth deviation from the linear trend can be seen in direct simulation of Equation~\ref{eq:EDSR_Hamiltonian} owing to higher-order terms, the origin of the numerous nonlinear features we observe is unclear. Furthermore, previous works in similar Si/SiGe devices have found Larmor frequency shifts of both signs that are not quadratic in driving amplitude \cite{Takeda_2018_freqshift, Watson_2018}, contrary to the theoretical expectation. This leads us to conclude that the model of Eq.~\ref{eq:EDSR_Hamiltonian} does not adequately capture all relevant features of the qubit physics.

Anharmonic models of the confinement potential $H_0$ have been used to explain nonlinear phenomena such as second-harmonic driving \cite{Scarlino_2015, Scarlino_2017} and even nonlinear Rabi scaling \cite{Khomitsky_2012_slowRabi,Tokura_2013_power}. However, with both valley splittings of the evaluated device measured to be in excess of \SI{150}{\micro\eV}, it is unclear why such an anharmonic confinement potential applies to this device. Furthermore, our numerical simulations of Eq.~\ref{eq:EDSR_Hamiltonian} with anharmonic orbital- and valley-like models fail to capture the breadth of nonlinear features we observe in experiment (see Appendix C).
%% End Model %%

\begin{figure}[t]
    \centering
    \includegraphics[width=\linewidth]{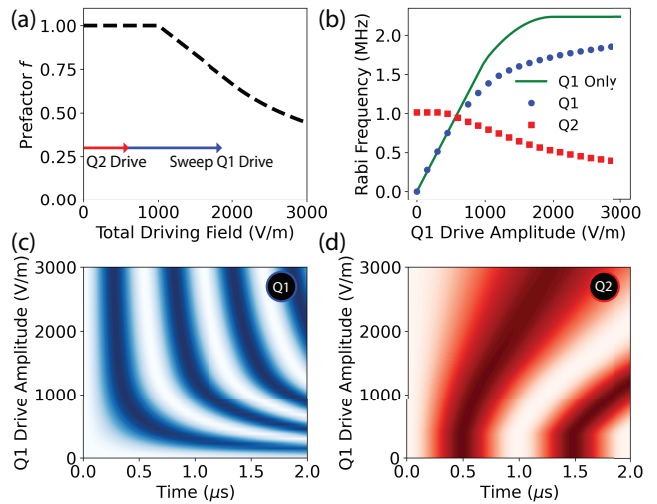}
    \caption{(a) Plot showing one possible instance of the phenomenological prefactor describing the nonlinearity in the EDSR mechanism. To illustrate the emergence of crosstalk, we set the Q2 electric drive to a constant amplitude, and manipulate the amplitude of the Q1 drive. The effective driving term in the Hamiltonian will be unique depending on the sum of both microwave drives. (b) The effect of two microwave drives on the two-spin system is numerically simulated with the nonlinear prefactor in (a). The solid green line gives the modified analytic Rabi frequency in the case that only Q1 is driven, while the discrete points are derived by fitting the numerically solved spin dynamics to a sinusoid. (c-d) The spin dynamics of Q1 and Q2 corresponding to the fits in (b). Light and dark regions indicate the probability of measuring a ground or excited state spin respectively. For simulation, we take $E_{Z,\mathrm{Q1}}=\SI{12.066}{\giga\hertz}$, $E_{Z,\mathrm{Q2}}=\SI{11.966}{\giga\hertz}$, $|\vec{b}_{SL}|=\SI{0.3}{\milli\tesla/\nano\meter}$, $a_0=\sqrt{\hbar/m^*\omega_0}=\SI{20}{\nano\meter}$, and $E_{ac,Q2}=\SI{600}{\volt/\meter}$.}
    \label{fig:4}
\end{figure}
%
%% Start Discussion %%
Based on the variety of nonlinearities observed from single-qubit measurements in Fig.~\ref{fig:2}, it is clear that at least the microwave power ($P_k$) and frequency ($\omega_k$) components are important contributing factors. We therefore focus on the time-dependent driving term of the EDSR Hamiltonian as the simplest source of the nonlinearity and identify two possible physical origins: electric drive distortion and microwave-induced artefacts.

If the signal amplitude at room temperature is not linearly related to the amplitude delivered to the device, then the origin of the nonlinearity may be trivially related to classical electronics or transmission lines. It is not possible for us to measure the electric field at the dot location without considering the electron spin as a sensor itself. Still, we have verified that the output of the IQ-modulated signal is linear with respect to the input. Beyond this element, there are known interference effects in the transmission lines but no active electronic components that are suspected to show nonlinear effects. Furthermore, the signal paths to the ``B'' and ``MW'' gates are separate from room temperature, and nonlinearity is present when either gate is used for driving. Conversely, applying a drive through the same gate gives a different nonlinear response depending on which qubit it addresses. This points to a microscopic origin of the nonlinearity, although the driving frequencies and orientations for the two qubits differ as well, making it difficult to completely rule out any origin of nonlinearity from a classical distortion of the driving signal. In Appendix A, we describe how nonlinearity and crosstalk were observed in a different experimental setup using a nominally identical device design. This reinforces the likelihood that the nonlinearity originates at the device and highlights that the nonlinear behaviour is not a peculiarity of a single experimental setup.

Second, we consider the possibility that a microwave drive could influence a quantum dot's confinement potential, and therefore its orbital structure, through heat-induced device strain or the activation of charge traps, for example. Although a true harmonic confinement potential is robust against small perturbations, the anharmonicity introduced by asymmetric confinement or valley-orbit hybridization may be sensitive to such changes \cite{Gamble_2013, Boross_2016, Hosseinkhani_2020_valley}. We therefore acknowledge the possibility that a nonlinear drive-dependent dipole element $r(E_\mathrm{tot}) = \langle VO_0(E_\mathrm{tot})|\hat{x}|VO_1(E_\mathrm{tot})\rangle$ may manifest in a way that is consistent with our observations. The plausibility of these hypotheses would need to be verified through more rigorous modelling.

Although the origin of the nonlinearity remains uncertain, we can nevertheless gain insight in the crosstalk effect by extending the model of Eq.~\ref{eq:EDSR_Hamiltonian} phenomenologically by including a prefactor $f(P_k,\omega_k)$ in the electric driving term such that $E'_{ac}(t)\hat{x}\rightarrow f(P_k,\omega_k)E'_{ac}(t)\hat{x}$. Following from the Rabi scalings measured in Fig.~\ref{fig:2}, we consider the prefactor to be dependent on the power $P_k\propto |E_{ac,k}|^2$ and frequency $\omega_k$ of all applied drives. To illustrate the consequences of the phenomenological model, consider the prefactor plotted in Fig.~\ref{fig:4}(a). As the total applied electric field increases, the effective driving amplitude no longer rises proportionally. To see the importance of this dependence, we numerically integrated the time-dependent Schr\"{o}dinger equation $i\hbar \dot{\psi} = H(t)\psi$ using Eq.~\ref{eq:EDSR_Hamiltonian} and the prefactor depicted in Fig.~\ref{fig:4}(a). A constant driving tone resonant with Q2 drives Rabi oscillations. As a Q1 driving tone is turned on, two notable crosstalk effects occur [Fig.~\ref{fig:4}(b-d)]. First, the Q1 Rabi frequency is substantially smaller than in the case where no Q2 drive is present. Second, the Q2 Rabi frequency decreases markedly as the Q1 Rabi frequency increases. The fitted Rabi frequencies in Fig.~\ref{fig:4}(b) behave analogously to the measured crosstalk effect presented in Fig.~\ref{fig:3}, and the same effect is obtained in the absence of a second electron.

For the near-term scaling of silicon spin qubit devices using micromagnet-based EDSR, the practical issues introduced here can be limited by ensuring the electric drive is oriented parallel to a sufficiently large transverse magnetic gradient. This ensures that a reasonably large $f_\mathrm{Rabi}$ is achieved at a sufficiently small magnitude of $E_{ac}$ that is within the perturbative limit. Our observations suggest multiplexing qubit control using a linear combination of driving signals is possible within this regime. However, the problem of non-linear Rabi scaling may persist at larger scales. Although we cannot provide a conclusive origin for this effect, we believe a more careful consideration of microwave propagation at the device will be fruitful.

\section{Conclusion}

In summary, we have presented experimental evidence of a strong nonlinearity in the fundamental resonance of a single-electron spin qubit controlled by EDSR with a synthetic spin-orbit coupling. To understand both the nonlinear Rabi frequency scaling and crosstalk effects that are observed, we have developed a simple phenomenological model whose accuracy and consequences may be probed through further experiments. The novel crosstalk mechanism introduced here poses important questions for the scalability of spin qubit devices relying on multiplexed single-qubit control.
%% End Discussion %%

% uncomment iffalse for word count
%\iffalse

\begin{acknowledgments}
We acknowledge useful discussions with Peihao Huang and members of the Vanderypen group. This work is supported by the Dutch Research Council, the European Union's Horizon 2020 research and innovation programme (QLSI grant 951852) and the Army Research Office (ARO) under grant number W911NF-17-1-0274. M.R. acknowledges support from the Netherlands Organization of Scientific Research (NWO) under Veni grant VI.Veni.212.223. Data and analysis scripts supporting this work are available at zenodo, https://doi.org/10.5281/zenodo.6473239.
\end{acknowledgments}
%\fi

\appendix
\section{Evidence of nonlinear Rabi scaling in a second device}

\begin{figure}[t]
    \centering
    \includegraphics[width=\linewidth]{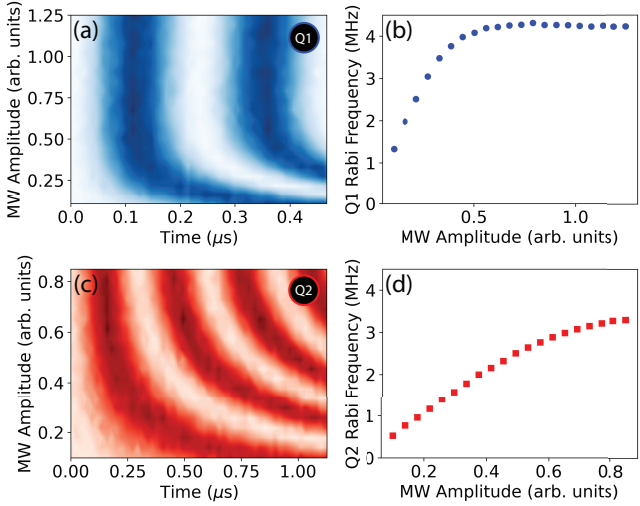}
    \caption{Rabi oscillations in Device B. Panels (a) and (c) plot the qubit dynamics as a function of microwave driving amplitude. Panels (b) and (d) show the fitted Rabi frequencies respectively.}
    \label{fig:A1}
\end{figure}

To provide further evidence that the nonlinear Rabi frequency scaling and crosstalk as discussed in the main text can be seen more generally, we include data collected from a second device - Device B - which is nominally identical with respect to the design in Fig.~\ref{fig:1}(a) - Device A - and is fabricated on the same purified $^{28}$Si/SiGe heterostructure. As such, we expect Device B to host quantum dots with a similar orbital confinement and micromagnet gradient as Device A. The Larmor frequencies of Q1 and Q2 of Device B were \SI{15.582}{\giga\hertz} and \SI{15.798}{\giga\hertz} respectively. Device B was cooled in an independent dilution refrigerator and controlled using different electronics than Device A. Rabi oscillations and fitted Rabi frequencies are included in Fig.~\ref{fig:A1}. The amplitude scale is independent of that used in the main text.

While both qubits reach the nonlinear regime at modest Rabi frequencies similar to Device A, Q1 of Device B illustrates a striking example of a very flat plateau. As with Device A, there is no immediate evidence of visibility loss or lower Rabi oscillation quality in the plateau region. However, the logical gate fidelities and $T_2^{Rabi}$ were not quantified. Crosstalk, of the kind described in the main text, was also observed in Device B, but not studied systematically like in the case of Device A. The observation of nonlinearity in a second device from an independent setup suggests that the origin of the phenomena can be attributed to the devices, and is less likely the result of a faulty component, for example.

\section{EDSR in a harmonic confinement potential}

\begin{figure}[t]
    \centering
    \includegraphics[width=\linewidth]{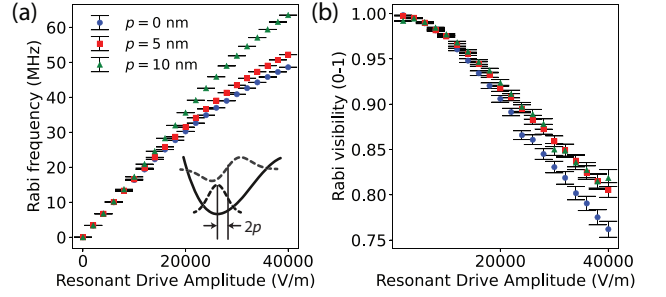}
    \caption{(a) Rabi frequency and (b) visibility of orbital-mediated EDSR where $\Delta_0=\SI{1}{\milli\eV}$, $r = 20/\sqrt{2}~\SI{}{\nano\meter}$, and $E_Z = \SI{60}{\micro\eV}$. As illustrated in the sketch, a larger length $p$ corresponds to a more skewed confinement potential.}
    \label{fig:A2}
\end{figure}

Here we summarize micromagnet-based EDSR with a harmonic confinement potential. We consider the following EDSR Hamiltonian:

\begin{eqnarray}
    \label{eq:EDSR_Hamiltonian_Harmonic}
    H(t) = \hbar\omega_0(\hat{a}^\dagger\hat{a}+\frac{1}{2}) + \tilde{E}_{ac}\sin(\omega t)(\hat{a}^\dagger + \hat{a})\nonumber\\
    - \frac{E_Z}{2}\sigma_z + \tilde{b}_{SL}(\hat{a}^\dagger+\hat{a})\vec{n}\cdot\vec{\sigma},
\end{eqnarray}

\noindent where the position operator $\hat{x} = \sqrt{\frac{\hbar}{2m^*\omega_0}}(\hat{a}^\dagger+\hat{a})$ in terms of the quantum harmonic ladder operators, $m^*=0.19m_e$ is the in-plane effective mass of the electron in the silicon quantum well, and the spin-dependent terms are as defined in the main text. For the subsequent analysis, the length scale is absorbed in the relevant energy scales such that $\tilde{b}_{SL} = \frac{1}{2}g\mu_B|\vec{b}_{SL}|\sqrt{\hbar/2m^*\omega_0}$ and $\tilde{E}_{ac} = eE_{ac}\sqrt{\hbar/2m^*\omega_0}$. By considering a typical orbital spacing of $\hbar\omega_0\approx$~\SI{1}{\milli\eV} corresponding to a Fock-Darwin radius of $a_0=\sqrt{\frac{\hbar}{m^*\omega_0}}\approx$~\SI{20}{\nano\meter}, an ac electric field amplitude of order less than $10^4$~\SI{}{\volt/\meter}, an external magnetic field of \SI{475}{\milli\tesla}, and a transverse magnetic field gradient of \SI{0.5}{\milli\tesla/\nano\meter}, the relevant terms correspond to the energy scales:

\begin{align*}
    \hbar\omega & \approx E_Z, \\
    \tilde{b}_{SL} & \approx \SI{1}{\micro\eV}, \\
    \tilde{E}_{ac} & \approx \SI{100}{\micro\eV}, \\
    E_Z & \approx \SI{50}{\micro\eV}. \\
\end{align*}

\noindent Therefore:
\begin{equation}
    \epsilon \approx \frac{\omega}{\omega_0} \approx \frac{\tilde{b}_{SL}}{\hbar\omega_0} \approx \frac{\tilde{E}_{ac}}{\hbar\omega_0} \approx \frac{E_Z}{\hbar\omega_0} \ll 1,
\end{equation}

\noindent and it is appropriate to treat all terms of order $\epsilon$ perturbatively with respect to the orbital energy scale. Following the approach of time-dependent Schrieffer-Wolff perturbation theory employed in \cite{Romhanyi_2015_EDSR}, we derive the effective spin Hamiltonian (ignoring elements proportional to the identity) up to fifth-order as $\tilde{H} = \sum_{n=1}^5 \tilde{H}^{(n)}$ where:

\begin{widetext}
\begin{align*}
    \tilde{H}^{(1)} & = \frac{-E_Z}{2}\sigma_z \\
    \tilde{H}^{(2)} & = -\frac{2\tilde{b}_{SL}\tilde{E}_{ac}\cos\theta\sin(\omega t)}{\hbar\omega_0}\sigma_y - \frac{2\tilde{b}_{SL}\tilde{E}_{ac}\sin\theta\sin(\omega t)}{\hbar\omega_0}\sigma_z \\
    \tilde{H}^{(3)} & = -\frac{E_Z\tilde{b}_{SL}^2\cos\theta\sin\theta}{\hbar^2\omega_0^2}\sigma_y + \frac{E_Z\tilde{b}_{SL}^2\cos^2\theta}{\hbar^2\omega_0^2}\sigma_z \\
%\end{align*}
%\begin{widetext}
%\begin{align*}
    \tilde{H}^{(4)} & = -\frac{\tilde{b}_{SL}\tilde{E}_{ac}\cos\theta(E_Z^2+\hbar^2\omega^2)\sin(\omega t)}{\hbar^3\omega_0^3}\sigma_y - \frac{\tilde{b}_{SL}\tilde{E}_{ac}\sin\theta\omega^2\sin(\omega t)}{\hbar\omega_0^3}\sigma_z \\
    \tilde{H}^{(5)} & = -\frac{E_Z\tilde{b}_{SL}^2\cos\theta\sin\theta(E_Z^2 + 2\tilde{E}_{ac}^2-\tilde{b}_{SL}^2-2\tilde{E}_{ac}^2\cos(2\omega t))}{\hbar^4\omega_0^4}\sigma_y + \frac{E_Z\tilde{b}_{SL}^2\cos^2\theta(E_Z^2-\tilde{b}_{SL}^2+4\tilde{E}_{ac}^2\sin^2(\omega t))}{\hbar^4\omega_0^4}\sigma_z.
\end{align*}
\end{widetext}

Expanding the Floquet Hamiltonian and carrying out another second-order Schrieffer-Wolff transformation \cite{Romhanyi_2015_EDSR} yields an on-resonance Rabi frequency of:

\begin{align}
    \label{eq:floquet_Rabi}
    \hbar\Omega_\mathrm{Rabi} & = \frac{2\tilde{b}_{SL}\tilde{E}_{ac}\cos\theta}{\hbar\omega_0}\left(1+\frac{E_Z^2}{\hbar^2\omega_0^2}\right)\\
    & = \frac{g\mu_Ba_0^2|\vec{b}_{SL}|\cos\theta eE_{ac}}{2\hbar\omega_0}\left(1+\frac{E_Z^2}{\hbar^2\omega_0^2}\right)
\end{align}

\noindent accurate to $E_Z\epsilon^4$. The drive-dependent frequency shift, analogous to the Bloch-Siegert shift in electron spin resonance, is given as:

\begin{align}
    \hbar\omega_\mathrm{BSS} & = -\frac{4E_Z\tilde{b}_{SL}^2\tilde{E}_{ac}^2\cos^2\theta}{\hbar^4\omega_0^4} \\
    & = -\frac{g^2\mu_B^2a_0^4E_Z|\vec{b}_{SL}|^2\cos^2\theta e^2E_{ac}^2}{4\hbar^4\omega_0^4}.
\end{align}

\noindent The sign of the shift is opposite what is expected from standard ESR. The reason for this is discussed in \cite{Romhanyi_2015_EDSR}. The resonance frequency shifts due to a non-linear Zeeman term and g-factor renormalization are calculated to be, respectively:

\begin{align}
    \hbar\omega_{nlz} & = -\frac{2E_Z^3\tilde{b}_{SL}^2\cos^2\theta}{\hbar^4\omega_0^4} \\
    \hbar\omega_g & = -\frac{2E_Z\tilde{b}_{SL}^2\cos^2\theta}{\hbar^2\omega_0^2}\left(1-\frac{\tilde{b}_{SL}^2}{\hbar^2\omega_0^2}\right).
\end{align}

\noindent Therefore, a harmonic confinement potential should yield the relations $f_\mathrm{Rabi}\propto E_{ac}$ and $\hbar\omega_\mathrm{BSS}\propto -E_{ac}^2$ for micromagnet-based EDSR. The perturbative regime used to derive these relations should be valid, for realistic parameters, at least to the order of $f_\mathrm{Rabi}=\SI{10}{\mega\hertz}$. It should be noted that nonlinear phenomena, such as second-harmonic driving, are permissible even with perfect harmonic confinement, as evidenced from the presence of longitudinal driving in $\tilde{H}^{(3)}$ and $\tilde{H}^{(5)}$. However, it is believed that nonlinearity originating from anharmonic confinement will be dominant in silicon quantum dots \cite{Scarlino_2015}.

\section{EDSR with anharmonic confinement}

\begin{figure}[t]
    \centering
    \includegraphics[width=\linewidth]{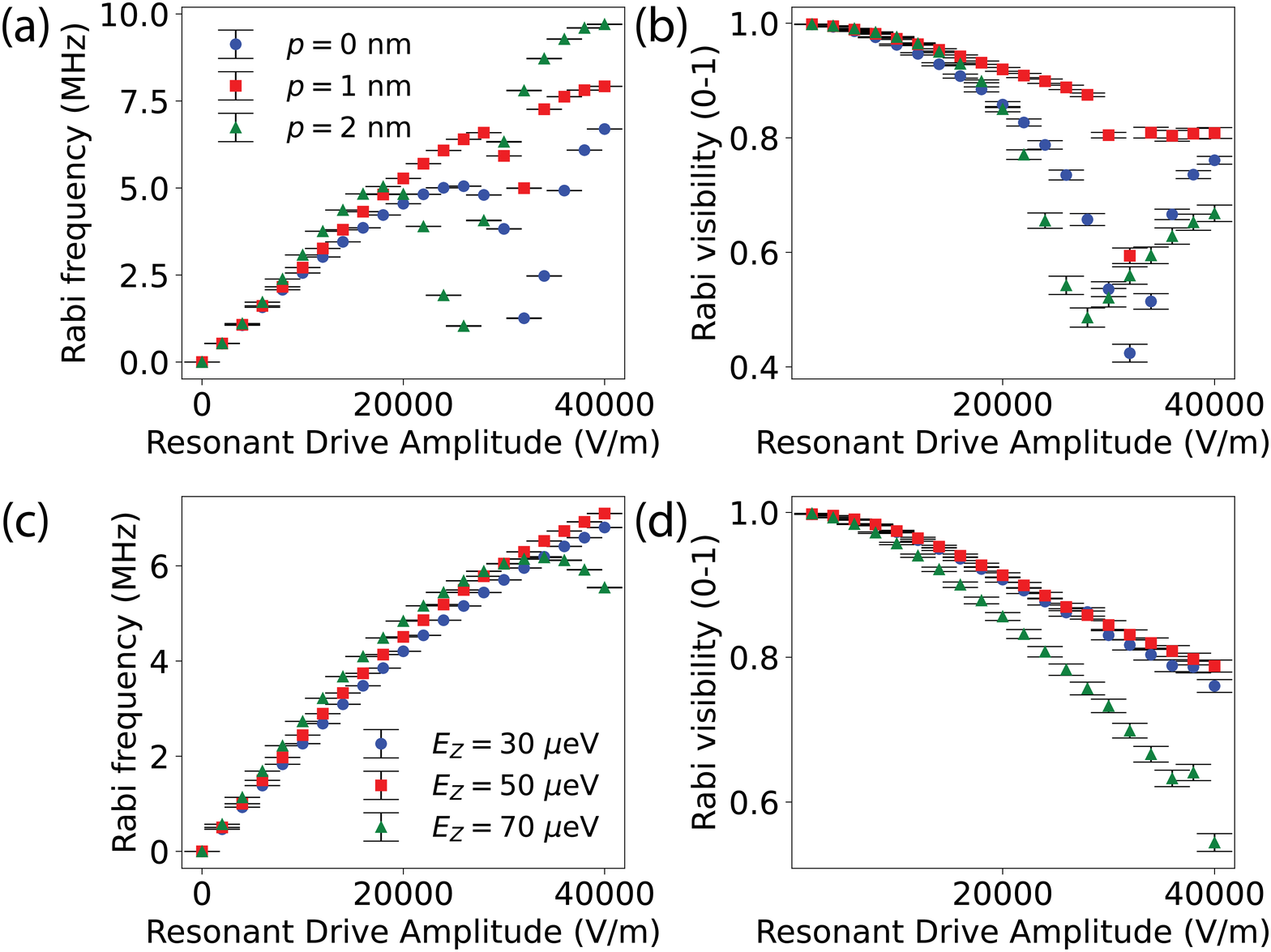}   
    \caption{(a) Rabi frequency and (b) visibility of valley-mediated EDSR where $\Delta_0=\SI{150}{\micro\eV}$, $r=\SI{2}{\nano\meter}$, and $E_Z = \SI{60}{\micro\eV}$. Different values of $p$ consider different spatial orientations of the excited valley state. (c) Rabi frequency and (d) visibility of valley-mediated EDSR considering different Zeeman splittings where $\Delta_0=\SI{150}{\micro\eV}$, $r=\SI{2}{\nano\meter}$, and $p = \SI{0}{\nano\meter}$.}
    \label{fig:A3}
\end{figure}

Here we show how EDSR in the presence of anharmonic confinement, either as a result of a nontrivial potential landscape or the presence of valley-orbit states, permits nonlinear phenomena. However, the nonlinear Rabi scaling found here does not seem to adequately account for the experimental results.

We consider a general two-level orbital subspace acted on by the set of Pauli operators $\{\tau_i\}$, which may describe two hybridized valley-orbit states in silicon, or the lowest two states of an anharmonic confinement potential. Since the micromagnet spin-orbit coupling energy scale is the smallest, we consider the dynamics of the driven orbital sector first. The driven orbital Hamiltonian is:

\begin{equation}
    H_0(t) = -\frac{\Delta_0}{2}\tau_z + E'_{ac}\sin(\omega t)\hat{x},
    \label{eq:LSZ_Hamiltonian_simple}
\end{equation}

\noindent where $\Delta_0$ denotes the energy splitting between the ground and excited states and $E'_{ac}=eE_{ac}$ is the scaled electric field as in the main text. The eigenstates of $H_0(t)$ when no drive is present, which we denote as $\ket{VO_0}$ and $\ket{VO_1}$, may in general contain both transverse and longitudinal elements, such that:

\begin{equation}
    \hat{x} = r\tau_x - p\tau_z,
\end{equation}

\noindent where $r = \bra{VO_0}\hat{x}\ket{VO_1} > 0$ and $2p = \bra{VO_1}\hat{x}\ket{VO_1} - \bra{VO_0}\hat{x}\ket{VO_0}$ are real parameters. The parameter $p$ quantifies the extent to which the orbital states have a different centre of mass, as would be the case in an asymmetric confinement potential (see the sketch in Fig.~\ref{fig:A2}). We transform the Hamiltonian by an angle $\pi/2-\theta$ about the $\tau_y$ axis, where $\sin\theta=\frac{p}{\sqrt{r^2+p^2}}$:

\begin{align}
    H_0'(t) & = \exp(i(\pi/2-\theta)\tau_y) H_0(t) \exp(-i(\pi/2-\theta)\tau_y) \\
    & = -\frac{\Delta}{2}\tilde{\tau}_x - \left(\frac{\epsilon+E_{ac}''\sin(\omega t)}{2}\right)\tilde{\tau}_z
    \label{eq:LSZ_Hamiltonian}
\end{align}

\noindent where $\Delta = \Delta_0\cos\theta$, $\epsilon = \Delta_0\sin\theta$, and $E''_{ac} = 2E'_{ac}\sqrt{r^2+p^2}$. Eq.~\ref{eq:LSZ_Hamiltonian} is the standard Landau-Zener-St\"uckelberg Hamiltonian \cite{Shevchenko_2010_LZS}. By moving into the rotating frame using the unitary transformation $\exp(-iE''_{ac}\cos(\omega t)\tilde{\tau}_z/2\omega)$ and applying the Jacobi-Anger expansion, one can distinguish between single-photon transition matrix elements $\Delta J_1(E_{ac}''/\hbar\omega)$ and two-photon transition matrix elements $\Delta J_2(E_{ac}''/\hbar\omega)$ where $J_n(x)$ is the n-th order Bessel function of the first kind. The latter mechanism corresponds to subharmonic driving, when the spin degree of freedom is included perturbatively. An analysis of Eq.~\ref{eq:LSZ_Hamiltonian} using the dressed-state formalism in the context of silicon-based EDSR is found in \cite{Scarlino_2017}.

To illustrate how the Rabi frequency scales in different parameter regimes, we consider the full Hamiltonian numerically for various $\Delta_0, r, p, E_Z$:

\begin{equation}
    \label{eq:sim_Hamiltonian}
    H(t) = -\frac{\Delta_0}{2}\tau_z + E'_{ac}\sin(\omega t)\hat{x} -\frac{E_Z}{2}\sigma_z + b'_{SL}\hat{x}\sigma_x,
\end{equation}

\noindent with all definitions the same as in the main text. In all cases, we choose an initial state in the ground valley-orbit and spin states, and we set $\hbar\omega = E_Z$ for all simulations, neglecting the small g-factor renormalization due to the micromagnet coupling for simplicity. The nominal transverse micromagnet gradient is simulated to be between 0, along the qubit axis, and \SI{0.7}{\milli\tesla/\nano\meter} along the orthogonal axis. Due to fabrication imperfections and qubit driving likely not taking place along the maximal gradient, we use a value of $|b_{SL}|=\SI{0.3}{\milli\tesla/\nano\meter}$ in simulations. The small longitudinal gradient can be predicted from the approximately \SI{100}{\mega\hertz} Zeeman difference between the qubit, with an estimated pitch of \SI{100}{\nano\meter}, as \SI{0.04}{\milli\tesla/\nano\meter} along the qubit axis. Therefore, we neglect any $\sigma_z$ coupling in the simulation.

In Fig.~\ref{fig:A2}, we simulate EDSR mediated by an orbital state with an estimated energy splitting of $\Delta_0=\SI{1}{\milli\eV}$ and a corresponding dipole transition element of $r=20/\sqrt{2}~\SI{}{\nano\meter}$. By changing the parameter $p$, we effectively model the influence of an asymmetric confinement potential, where the excited state has a shifted centre of mass. Such a skew has negligible influence for a weakly driven spin. Notably, for larger drives ($f_\mathrm{Rabi}\gg \SI{10}{\mega\hertz}$) the Rabi frequency deviates below the linear trend predicted by Eq.~\ref{eq:floquet_Rabi} and the visibility decreases due to residual couplings to spin-orbit states outside of the qubit subspace. Such effects have been observed in \cite{Takeda_2016,Yoneda_2017}, and may be exacerbated by microwave heating which is not included in our simulations. For considering the phenomenology discussed in the main text, we use this model when $p = \SI{0}{\nano\meter}$ and add the prefactor $f(P_k,\omega_k)$ to Eq.~\ref{eq:EDSR_Hamiltonian_Harmonic}.

In Fig.~\ref{fig:A3}(a-b), we probe EDSR when mediated via a valley state. Magnetospectroscopic measurements of the device in our experiment show valley splittings of \SI{180}{\micro\eV} and \SI{160}{\micro\eV} for Q1 and Q2 respectively. We note that valley splittings of this magnitude would suggest that there is a relatively small degree of hybridization between orbital and valley degrees of freedom due to interface defects. For our simulations, we select a valley splitting of $\Delta_0=\SI{150}{\micro\eV}$ and a modest dipole transition element of $r=\SI{2}{\nano\meter}$. An interesting feature appears at the particular Zeeman splitting of $E_Z=\SI{60}{\micro\eV}$, where both a dip in the Rabi frequency and visibility are found at larger driving amplitudes. The precise driving amplitude where this dip occurs depends on the spatial nature of the valley-like states. We have verified that such a dip can also result in a crosstalk effect, though it qualitatively does not match that observed in experiment.

In Fig.~\ref{fig:A3}(c-d), we repeat EDSR simulations with a valley-like state with $r=\SI{2}{\nano\meter}$ and $p=\SI{0}{\nano\meter}$ for different Zeeman splittings similar to those used in experiment. In contrast to orbit-mediated EDSR, there is a notable dependence of the Rabi frequency on $E_Z$. However, no nonlinearity like that seen in experiment is observed.

%\end{document} % uncomment for word count

% Run with the below command first, then copy/paste the output.bbl file to put references into a single .tex file.
%\bibliography{referencesBU}

%\iffalse

%% End Bibliography %%
%\fi

\end{document}